\documentclass{article}

\usepackage{microtype}
\usepackage{graphicx}
\usepackage[table]{xcolor}
\usepackage{booktabs} 
\usepackage{xcolor}
\usepackage{multirow}
\usepackage{subcaption}
\usepackage{diagbox}
\usepackage{makecell}
\usepackage{amsmath}
\usepackage{diagbox}
\usepackage{tcolorbox}
\usepackage{listings}
\usepackage[flushleft]{threeparttable}

\usepackage{hyperref}

\usepackage[accepted]{mlsys2025}

\mlsystitlerunning{The Hidden Bloat in Machine Learning Systems}

\colorlet{punct}{red!60!black}
\definecolor{background}{HTML}{EEEEEE}
\definecolor{delim}{RGB}{20,105,176}
\colorlet{numb}{magenta!60!black}

\lstdefinelanguage{asm}{
basicstyle=\normalfont\ttfamily,
numbers=none,
numberstyle=\scriptsize,
stepnumber=1,
numbersep=8pt,
showstringspaces=false,
breaklines=true,
frame=lines,
backgroundcolor=\color{background},
}

\begin{document}

\twocolumn[
\mlsystitle{The Hidden Bloat in Machine Learning Systems}




\begin{mlsysauthorlist}
\mlsysauthor{Huaifeng Zhang}{to}
\mlsysauthor{Ahmed Ali-Eldin}{to}
\end{mlsysauthorlist}

\mlsysaffiliation{to}{Chalmers University of Technology, Sweden}
\mlsysaffiliation{to}{Chalmers University of Technology, Sweden}

\mlsyscorrespondingauthor{Huaifeng Zhang}{huaifeng@chalmers.se}
\mlsyscorrespondingauthor{Ahmed Ali-Eldin}{ahmed.hassan@chalmers.se}

\mlsyskeywords{Machine Learning Systems, Software Bloat, Machine Learning Frameworks, Software Debloating, Software Engineering}

\vskip 0.3in

\begin{abstract}
Software bloat refers to code and features that are not used by a software during runtime.  For Machine Learning (ML) systems, bloat is a major contributor to their technical debt, leading to decreased performance and resource wastage.  
In this work, we present Negativa-ML, a novel tool to identify and remove bloat in ML frameworks by analyzing their shared libraries.
Our approach includes novel techniques to detect and locate unnecessary code within GPU code - a key area overlooked by existing research.
We evaluate Negativa-ML using four popular ML frameworks across ten workloads over 300 shared libraries.
Our results demonstrate that ML frameworks are highly bloated on both the GPU and CPU code side, with GPU code being a primary source of bloat within ML frameworks.
On average, Negativa-ML reduces the GPU code size by up to 75\% and the CPU code by up to 72\%, resulting in total file size reductions of up to 55\%.
Through debloating, we achieve reductions in peak CPU memory usage, peak GPU memory usage, and execution time by up to 74.6\%, 69.6\%, and 44.6\%, respectively.

\end{abstract}
]



\printAffiliationsAndNotice{}  

\section{Introduction}

From personalized recommendations~\cite{ko2022survey}, to healthcare diagnostics~\cite{bhardwaj2017study}, and autonomous vehicles~\cite{parekh2022review}, ML is revolutionizing nearly every industrial sector. Besides
Large Language Models (LLMs) which enabled natural language understanding and content generation~\cite{openai2024gpt4technicalreport,touvron2023llamaopenefficientfoundation}, smaller models are also now widely deployed in many use-cases, from robots and autonomous vehicles, to cameras and mobile phones.
As these models grow in numbers, size, and complexity, managing and optimizing ML systems has become increasingly challenging.
LLMs often contain billions of parameters, requiring substantial computational resources and vast amounts of training data. Smaller models on the other hand, are typically deployed in resource-constrained environments.
To manage this complexity, substantial technical overhead is introduced to ML systems, exacerbating the issue of software bloat within these systems~\cite{zhang2024machine}.

Software bloat refers to code that is unnecessary for a program during runtime, typically caused by extraneous functions, libraries, or features that do not contribute to the core functionality.
Software bloat can cause a range of issues, including decreased performance, increased resource usage, and security vulnerabilities.
While bloat can affect any type of software, ML systems have a special capacity for incurring it, as they have all the maintenance problems of traditional code plus an additional set of ML-specific issues, such as boundary erosion, data dependencies, and so on~\cite{sculley2015hidden}.
As ML models grow in scale and complexity, the bloat in ML systems increases, leading to additional inefficiencies that hinder runtime performance and increase the cost of deployment, particularly in resource-constrained environments~\cite{zhang2024machine,jiang2020mnn}.

At the heart of ML systems are ML frameworks, such as TensorFlow~\cite{abadi2016tensorflow} and PyTorch~\cite{paszke2019pytorch}.
These frameworks provide the essential libraries and tools for model training and inference.
ML frameworks are typically developed using multiple programming languages with
C++ and CUDA used to implement the core functionalities in order to maximize performance. Both C++ and CUDA code are compiled into shared libraries.
Python acts as the frontend, wrapping these core functionalities and enhancing the frameworks' usability for developers.
However, with the ease of use and flexibility that these frameworks provide, they also introduce \textit{framework tax}~\cite{fernandez2023framework} -
these frameworks can introduce performance overheads and diminish the benefits of new hardware and model architecture advancements.
Furthermore, integrating these frameworks also leads to large binary sizes and brings unnecessary overhead for smaller GPUs~\cite{jiang2020mnn}.

In this paper, we aim to identify and reduce bloat in ML frameworks by debloating, i.e., removing the bloat, in ML shared libraries.
Shared libraries encapsulate the core functionalities of ML frameworks and constitute most of the size of ML framework. These libraries can be hundreds of megabytes to a few gigabytes in size.
Debloating ML shared libraries involves the following challenges:
\begin{itemize}
    \item ML frameworks rely on some proprietary libraries, such as cuDNN~\cite{nvidiaCUDADeep} and cuBLAS~\cite{nvidiaCuBLAS}. These libraries are not open-source, making it impossible to perform source code analysis.
    \item ML frameworks contain code that runs on both the CPU and GPU. GPU code is overlooked by existing research and no method measure and reduce bloat in it.
    \item Although CPU code has a well-defined structure and the bloat in it has been previously studied~\cite{brown2024broad}, the structure of GPU code is not publicly available, making it difficult to analyze and debloat.
\end{itemize}

To address these challenges, we propose \textit{Negativa-ML}~\footnote{Code will be available at: \url{https://github.com/negativa-ai/negativa-ml}}, a tool to identify and remove bloat in both CPU and GPU code within ML shared libraries.
Leveraging insights into how ML systems execute ML workloads on CPU and GPU, we propose a novel approach to detect the GPU code used by an ML workload with low performance overhead. Subsequently, we locate the file ranges occupied by this GPU code within ML shared libraries.
This involves a deep understanding of how GPU code is compiled and how they are organized within a shared library, which is challenging because GPU code  structure lacks an open specification.
Finally,  we utilize  a debloating tool that we have previously developed~\cite{zhang2025hiddenbloatmachinelearning} to remove bloat from ML shared libraries according to the file ranges located.
Our contributions are as follows:
\begin{itemize}
    \item We propose an approach to detect GPU code used by ML workloads with low performance overhead.
    \item We analyze the structure of GPU code in ML shared libraries and propose a method to locate the file ranges of used GPU code.
    \item We extend a debloating tool that we have developed~\cite{zhang2025hiddenbloatmachinelearning} to remove bloat in both CPU and GPU code in ML shared libraries.
    \item We evaluate Negativa-ML on four ML frameworks across ten ML workloads with three models and over 300 shared libraries and perform in-depth analysis of the results.
          Our evaluation shows the extent of bloat in ML frameworks, their causes and the overhead the bloat incurs.
\end{itemize}
Our evaluation shows that ML frameworks are highly bloated, with 72\% of CPU code and 75\% of GPU code being unnecessary for target ML workloads.
In addition, 10\% of the shared libraries account for up to 90\% of the total bloat in the ML frameworks.
This bloat not only increases storage overhead but also degrades runtime performance with increased CPU memory usage, GPU memory usage, and longer execution times.

\section{Background and Related Work}
This section provides background about ML frameworks, ML shared libraries, software bloat and debloating.

\subsection{ML Frameworks}

ML frameworks are software libraries that provide essential tools for building, training, and deploying ML models.
Some ML frameworks are general-purpose and designed to support both training and inference for a wide variety of ML models.
While others are optimized for specific use cases, such as LLM inference.
Two of the most popular general-purpose ML frameworks are TensorFlow~\cite{abadi2016tensorflow} and PyTorch~\cite{paszke2019pytorch}, both of which are widely adopted in industry and academia.
Although the two frameworks target various ML models, 
the rise of LLMs has introduced new requirements for ML frameworks, such as efficient KV cache management~\cite{kwon2023efficient,sinha2024llmops}.
LLM inference — which is highly latency-sensitive and resource-demanding — is particularly challenging.
To meet these demands, new frameworks specifically designed for LLM inference have been proposed~\cite{2023lmdeploy,kwon2023efficient,wolf-etal-2020-transformers, huggingfaceTextGeneration}.

The core functionalities of ML frameworks are packaged as shared libraries.
A shared library is a file that contains machine code that can be shared among different programs.
Shared libraries account for the majority of the total size of ML frameworks.
For example, in PyTorch and TensorFlow, shared libraries constitute 93\% and 75\% of the total framework size, respectively.

The standard file format for shared libraries is the Executable and Linkable Format (ELF)~\cite{oracleFileFormat},
which organizes a file into various sections, such as \texttt{.text} and \texttt{.data}.
A generic shared library only contains code running on CPU in the \texttt{.text} section.
However, ML shared libraries are unique in that they also contain the code designed to run on GPUs.
This GPU code is usually included in another section called \texttt{.nv\_fatbin} in an ML shared library.

\begin{figure}[ht]
    \centering
    \includegraphics[width=0.33\textwidth]{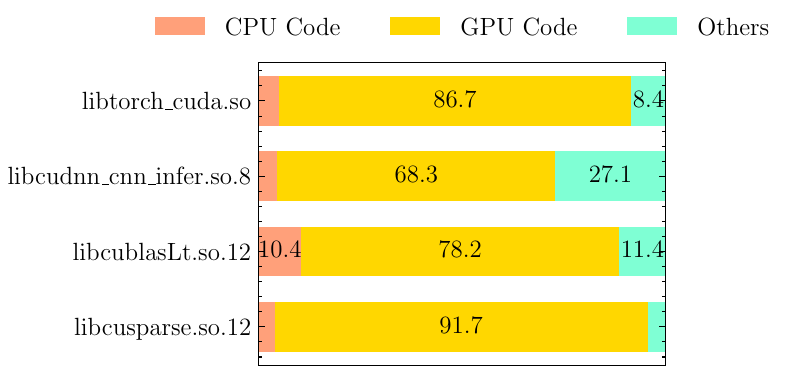}
    \caption{Distribution of CPU code and GPU code in the top 4 largest shared libraries in PyTorch.}
    \label{fig:orig-size-dist}
    \vspace{-2em}
\end{figure}

GPU code---also known as device code---significantly increases the size of ML shared libraries compared to traditional shared libraries.
The core shared library in PyTorch, \texttt{torch.so}, is 881 MB for the GPU version and 482 MB for the CPU version, nearly double the size.
Furthermore, GPU code makes up a substantial portion of these libraries.
Analyzing the top four largest shared libraries in PyTorch, GPU code accounts for between 68\% to more than 91\% of the size of each shared library, as shown in Figure~\ref{fig:orig-size-dist}.
In contrast, CPU code---also known as host code---represents only a small fraction of these libraries.
Despite GPU code forming the majority of ML shared libraries, no existing research has investigated bloat within GPU code.

\subsection{Software Bloat}
Software bloat is mainly caused by unnecessary code in a software, which can be categorized into two types~\cite{brown2024broad}:
Type I bloat is universally unnecessary, for example, dead and unreachable code.
Type II bloat is end-use dependent, code may be or may not be Type II bloat depending on the use case, for example, unnecessary program features can be used by other programs that use the shared library.
Software bloat exists across the entire modern software stack, from operating systems~\cite{quach2017multi,kuo2020set}, shared libraries~\cite{ziegler2019honey,quach2018dpiecewise,biswas2021ancile}, to executables~\cite{navas2023occamv2,ahmad2021trimmer,qian2019razor} and even containerized applications~\cite{rastogi2017cimplifier,zhang2024machine}.
Bloat causes software to grow in size and complexity over time, leading to performance degradation, increased memory consumption, longer startup times, and increased security vulnerabilities, without providing any benefits.
Recently, bloat in ML systems has attracted increasing attention.
Sculley et al.~\cite{sculley2015hidden} identify that ML systems are easy to accumulate hidden technical debt due to a set of ML-specific issues.
Zhang et al.~\cite{zhang2024machine} show that containerized ML applications are significantly bloated, wasting storage resources and network bandwidth, increasing their attack surface, and slowing down their deployment process.

\subsection{Software Debloating}

Software debloating is the process of removing unnecessary code from software to improve efficiency and reduce resource consumption.
Many debloating tools have been developed for traditional software, which can be categorized by their debloating targets: source code, binary code, and containerized applications.
Source code debloating tools eliminate unused code directly from the source, such as dead or unreachable code, based on specific usage scenarios~\cite{brown2019carve,ye2021jslim,azad2019less}.
Binary debloating tools operate on software binaries, including shared libraries and executables, to remove unnecessary functions or instructions~\cite{qian2019razor,ahmad2021trimmer,ziegler2019honey,agadakos2019nibbler}.
Container debloating tools target containerized applications, removing unnecessary files in container images~\cite{rastogi2017cimplifier,zhang2023blafsbloatawarefile}.
Most of these tools remove code that is not used by the specific workload, i.e., Type II bloat~\cite{qian2019razor,ahmad2021trimmer,brown2019carve,rastogi2017cimplifier,azad2019less}.
While debloating traditional software has been extensively studied, to the best of our knowledge, all existing work focus only on traditional applications where code runs only on CPUs. 
GPU code, which constitutes a significant portion of ML frameworks, remains unstudied.

One major issue with debloating tools that focus on binary debloating is that they are generally unreliable~\cite{brown2024broad}. To solve their reliability issues, we have recently developed Negativa, a debloating tools that only debloat CPU code~\cite{zhang2025hiddenbloatmachinelearning}.
Negativa demonstrates effectiveness in debloating shared libraries,  profiling workloads with the target shared libraries, then removing any code not used by the workload.
Negativa’s debloating process involves three phases: detection, location, and compaction.
In the detection phase, it identifies CPU functions used by the workload;
in the location phase, it locates the used CPU functions within the shared library;
and in the compaction phase, it removes unused functions from the shared library and only keeps the used ones.

In this work, we extend Negativa, introducing Negativa-ML, a technique that we have developed to enable GPU code debloating. In addition, we build on Negativa's capabilities to also debloat CPU code in ML shared libraries. Building on Negativa's three-phase debloating process, we introduce novel approaches for detecting and locating used GPU code in the detection and location phases. 
Negativa’s compaction phase is then reused to remove unused GPU code.

\section{Methodology}
\begin{figure}
    \centering
    \includegraphics[width=0.22\textwidth]{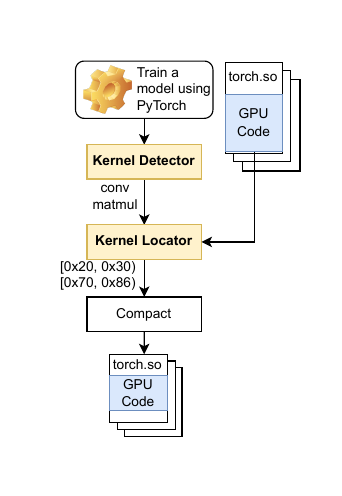}
    \caption{Overview of Negativa-ML. The components proposed in this work are highlighted in yellow rectangles.}
    \label{fig:debloat_overview}
    \vspace{-2em}
\end{figure}

Figure \ref{fig:debloat_overview} provides an overview of Negativa-ML.
Negativa-ML is composed of two components: the kernel detector and the kernel locator.
During the execution of an ML workload, many shared libraries of the ML framework are utilized.
The kernel detector monitors the kernel execution of the workload and records the names of the kernels used.
The kernel locator analyzes the shared libraries, extracting GPU code within these libraries and finding the file ranges of the used kernels within the shared library.
These ranges are subsequently passed to Negativa's compaction module for debloating.
Finally, a debloated ML shared library with reduced CPU code and GPU code is generated.

\subsection{Kernel Detector}
The kernel detector is responsible for detecting the kernels used by the ML workload.
Although existing tools like Nsight Systems (NSys)~\cite{nvidiaNVIDIANsys} can profile GPU code,  they are designed primarily to profile and debug GPU performance runtime rather than for kernel detection.
These tools, to provide comprehensive information, usually impose a high performance overhead on the target applications by recording data each time a kernel is called.
However, for kernel detection, we are only interested in determining whether a kernel has been used, without needing repetitive call data, which results in unnecessary overhead.
To offer a more efficient solution, we propose a novel, lightweight approach that specifically detects kernels used within ML shared libraries, achieving low performance overhead.
Since kernels execute on GPUs, an intuitive approach would be to monitor GPU execution directly.
However, this approach is not feasible because the code execution on GPUs is not directly accessible.

Our approach leverages insights into how ML systems execute ML workloads on both the CPU and the GPU;
Considering the interaction between CPU and GPUs, first, the CPU launches a kernel(s), then the \textit{CPU-launching kernel} may or maybe not launch other kernels.
A kernel launched from another kernel is called \textit{GPU-launching kernel}.
These kernels form a \textit{kernel call graph}, where the start of the graph is the CPU-launching kernel.
The kernel detector only detects CPU-launching kernels.
By monitoring the CPU code, we can identify kernels launched from the CPU and subsequently consider them as ``used''.
This allows us to detect used kernels without the need for GPU-level monitoring.

To launch a kernel from the CPU, the CUDA driver must call the CPU function \texttt{cuModuleGetFunction} first.
The function takes the name of the kernel to be launched as one of its inputs.
It returns a function handle, which is used to execute the kernel.
Moreover, \texttt{cuModuleGetFunction} is only called once for each kernel, no matter how many times the kernel is executed, making it ideally suited for our used kernel detection.
Inspired by this observation, we implement a hook to the \texttt{cuModuleGetFunction} using Nvidia CUPTI API~\cite{nvidiaCupti}.
This hook records the kernel names passed to the function, which are then considered as used kernels.
The kernel detector intercepts each CPU-launching kernel only once and does not intercept GPU-launching kernels.
Consequently, the performance overhead is lower than profiling tools like NSys.

\begin{figure}[ht]
    \begin{minipage}[b]{0.238\textwidth}
        \centering
        \includegraphics[width=1\textwidth]{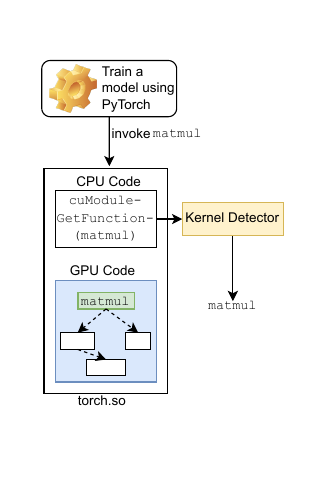}
        \caption{\small{Kernel detector workflow.}}
        \label{fig:kernel-detector}
    \end{minipage}
    \hfil
    \begin{minipage}[b]{0.238\textwidth}
        \centering
        \includegraphics[width=1\textwidth]{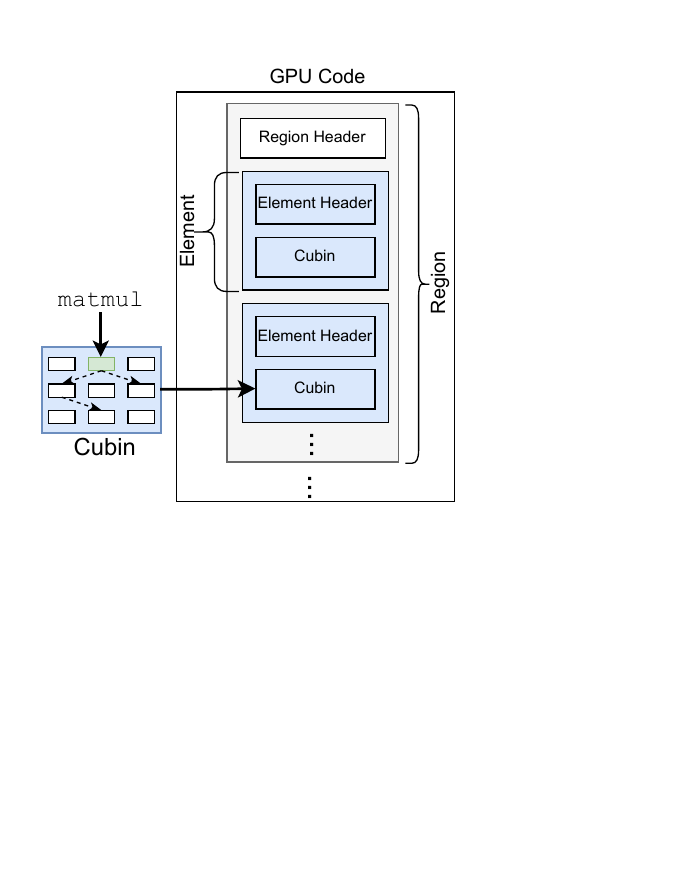}
        \caption{\small{Structure of GPU code.}}
        \label{fig:organiztion_GPU_code}
    \end{minipage}
    \vspace{-2em}
\end{figure}

Figure~\ref{fig:kernel-detector} illustrates an example of the kernel detector workflow.
When an ML workload invokes the \texttt{matmul} kernel, it first calls the \texttt{cuModuleGetFunction}, which resides in the CPU code, to launch the kernel.
The kernel detector intercepts the call to \texttt{cuModuleGetFunction} and records the kernel name, marked as a used kernel, which is \texttt{matmul} in this example.
This kernel may launch other kernels in the GPU code and form a kernel call graph.
However, these GPU-launching kernels are not detected by the kernel detector.

The kernel detector outputs a list of names of CPU-launching kernels.
These kernels are considered as used kernels and are passed to the kernel locator for further processing.
GPU-launching kernels are also handled by the kernel locator, as we show in the next section.

\subsection{Kernel Locator}
The kernel locator locates the file ranges (start and end file offset) of the used kernels in the ML shared library.
It identifies a list of file ranges that need to be retained in the shared library.
The difficulty in locating kernels in the GPU code is that there is no public specification for the structure of the GPU code.
Finding the exact location of used kernels requires in-depth analysis of the GPU code, which can be overfitted to a specific version of the CUDA toolkit and error-prone.
In addition, because GPU-launching kernels are not detected by the kernel detector, the kernel locator must also handle these kernels appropriately to avoid mistakenly removing them.

To address these issues, we propose an approach to \textit{approximately} locate the used kernels in the GPU code.
Instead of finding the exact location of kernels, we find the location of the \textit{cubins} that contain the kernels.
A cubin is a CUDA binary file that contains kernel code.
If a kernel is launched by another kernel, the two kernels are compiled into the same cubin~\cite{nvidiaNVIDIACUDA}.
Based on this observation, we can deduce that if a cubin contains a CPU-launching kernel, it also contains all the kernels in the kernel call graph starting from the CPU-launching kernel, including those GPU-launching kernels.
Therefore, retaining a cubin that contains a CPU-launching kernel also retains all the kernels in the kernel call graph starting from the CPU-launching kernel, including those GPU-launching kernels as well.

To find whether a cubin contains a CPU-launching kernel, we use the \texttt{cuobjdump} tool ~\cite{nvidiaCUDABinary} to extract a list of cubin files from the shared library.
For each cubin, we extract the kernels included in it using \texttt{cuobjdump}.
If a cubin contains a used CPU-launching kernel, the whole cubin needs to be retained.
In doing so, we also retain the GPU-launching kernels in the cubin.
The next step is to locate the cubin in the shared library file, i.e., find the file range occupied by the cubin.
This involves understanding the structure of the GPU code in the shared library.
As shown in Figure \ref{fig:organiztion_GPU_code}, the GPU code in a shared library is organized as a list of \textit{regions}.
Each region includes a region header and a list of \textit{elements}.
Each element includes an element header and a cubin.
A cubin extracted by \texttt{cuobjdump} has an index starting from one in its file name.
This index is equal to the index of the element containing the cubin in the shared library.
Using this index, we can map the cubin to the corresponding element in the shared library.
In doing so, we can locate the file range occupied by the cubin.

To maintain the integrity of the shared library, the kernel locator retains or removes the whole element containing a cubin.
Finally, the kernel locator uses the following criteria to determine whether to retain an element;
The element header has a field called \textit{compute-capability}, which shows the GPU architecture the cubin is compiled for.
We find that only the elements that match the GPU architecture can be loaded into the GPU memory.
Therefore, if an element matches the GPU architecture which the ML workload is running on \textit{and} contains a cubin that has used CPU-launching kernels, then we retain the element.

Figure~\ref{fig:organiztion_GPU_code} also shows an example of the relationship between kernels, cubins and elements in the shared library.
The used kernel \texttt{matmul}, which is detected by the kernel detector, is contained in a cubin as shown in the figure.
The kernel may launch other kernels, forming a kernel call graph.
All the kernels in the kernel call graph are contained in the same cubin.
The cubin is in turn contained in an element in the shared library.
By retaining the whole element, we ensure that all the kernels in the kernel call graph are retained, including the GPU-launching kernels in the call graph.

\noindent\textbf{Compaction:} The file ranges occupied by elements that meet the criteria are retained in the ML shared library, while the rest are removed.
This process is handled by Negativa's compaction phase. In this phase, the unused files are zeroed out. Negativa then maps
the file offsets to their original memory addresses where the
original shared library is loaded into memory to retain memory address validity. More details on the compaction process can be found in ~\cite{zhang2025hiddenbloatmachinelearning}.

\section{Experiments}

\begin{table*}[ht]
    \centering
    \scalebox{0.95}{
        \begin{threeparttable}
            \caption{Details of evaluated ML frameworks and ML workloads.}
            \label{tab:workloads}

            \begin{tabular}{llllrr}
                \toprule
                Model                                 & Framework                          & Operation & DataSet                                         & Batch Size & Epochs \\
                \midrule
                \multirow{4}{*}{\texttt{MobileNetV2}} & \multirow{2}{*}{PyTorch:2.3.1}     & Train     & CIFAR10~\cite{krizhevsky2009learning} Train Set & 16         & 3      \\
                                                      &                                    & Inference & CIFAR10  Test Set\tnote{1}                      & 4          & -      \\
                \cmidrule{2-6}
                                                      & \multirow{2}{*}{TensorFlow:2.16.2} & Train     & CIFAR10 Train Set                               & 16         & 3      \\
                                                      &                                    & Inference & CIFAR10   Test Set\tnote{1}                     & 4          & -      \\
                \midrule
                \multirow{4}{*}{\texttt{Transformer}} & \multirow{2}{*}{PyTorch:2.3.1}     & Train     & Multi30k~\cite{W16-3210}     Train Set          & 128        & 3      \\
                                                      &                                    & Inference & Multi30k    Test Set\tnote{1}                   & 32         & -      \\
                \cmidrule{2-6}
                                                      & \multirow{2}{*}{TensorFlow:2.16.2} & Train     & WMT14~\cite{bojar-EtAl:2014:W14-33}   Train Set & 128        & 1      \\
                                                      &                                    & Inference & WMT14   Test Set \tnote{1}                      & 32         & -      \\
                \midrule
                \multirow{2}{*}{\texttt{Llama2}}      & vLLM:0.6.3                         & Inference & Manual Input                                    & 1          & -      \\
                \cmidrule{2-6}
                                                      & Transformers:4.42.3                & Inference & Manual Input                                    & 1          & -      \\
                \bottomrule
            \end{tabular}
            \begin{tablenotes}
                \item[1] Only one batch from test set is used.
            \end{tablenotes}
        \end{threeparttable}
    }
\end{table*}

We evaluate Negativa-ML's bloat removal with four ML frameworks: two general-purpose ML frameworks, PyTorch and TensorFlow for their wide usage;
and two LLM inference frameworks, vLLM~\cite{kwon2023efficient} and Transformer{\textit{\textbf{s}}}~\cite{wolf-etal-2020-transformers} for their state-of-the-art performance.
Using these frameworks, we run various ML workloads with both training or inference of different popular models, to identify unnecessary code with respect to \textit{each} workload.

Table~\ref{tab:workloads} shows the details of the workloads.
For the ML models, we choose three models of different sizes:
A small model, \texttt{MobileNetV2}~\cite{sandler2018mobilenetv2}, which is a computer vision model of 4.3M parameters;
A medium model, \texttt{Transformer}~\cite{vaswani2017attention}, which is a natural language processing model of 65M parameters;
And a large model, \texttt{Llama-2-7b-chat-hf} (\texttt{LLama2} for brevity)~\cite{touvron2023llama}, which is a large language model of 7B parameters.
In total, 10 workloads were executed using the four ML frameworks.
We did not train the models to convergence, as our primary goal is to evaluate the bloat in the frameworks rather than fully train the models.
Since training mainly involves repeated iterations, training a few epochs is sufficient to obtain representative results.

All the workloads in Table~\ref{tab:workloads} were run on an AWS instance with 16 CPUs, 64 GB of memory, and an NVIDIA T4 GPU.
For the vLLM and Transformers frameworks, we also evaluate nine more LLM models using 8$\times$A100 GPUs, to further evaluate the bloat in a different hardware setup.


\subsection{Overview of Bloat in ML Frameworks}\label{sec:overview}
\begin{table*}[]
    \centering
    \caption{Total file size, CPU code, GPU code and their reductions of all shared libraries in each ML framework. The table shows the original value of a metric and the reduction in percentage of the metric in parentheses. K=1,000.}
    \label{tab:static-reduction}
    \scalebox{0.96}{
        \begin{tabular}{lllrrrrrr}
            \toprule
            \multirow{2}{*}{Model}                & \multirow{2}{*}{Framework}  & \multirow{2}{*}{Operation} & \multirow{2}{*}{\#Lib.} & \multirow{2}{*}{\makecell{Total File                                                     \\ Size/MB}}                      & \multicolumn{2}{c}{CPU Code} & \multicolumn{2}{c}{GPU Code}                        \\ \cmidrule(l{6pt}r{6pt}){6-7}\cmidrule(l{6pt}r{6pt}){8-9}
                                                  &                             &                            &                         &                                      & Size/MB  & \#Functions & Size/MB    & \#Elements  \\ \hline
            \multirow{4}{*}{\texttt{MobileNetV2}} & \multirow{2}{*}{PyTorch}    & Train                      & 113                     & 3,762 (55)                           & 557 (68) & 616K (93)   & 2,279 (75) & 14,062 (98) \\
                                                  &                             & Inference                  & 111                     & 3,569 (55)                           & 545 (70) & 616K (93)   & 2,103 (75) & 12,035 (98) \\ \cmidrule{2-9}
                                                  & \multirow{2}{*}{TensorFlow} & Train                      & 253                     & 3,274 (48)                           & 598 (46) & 984K (65)   & 1,774 (73) & 15,081 (99) \\
                                                  &                             & Inference                  & 251                     & 3,087 (47)                           & 586 (48) & 984K (65)   & 1,603 (72) & 13,056 (99) \\ \midrule
            \multirow{4}{*}{\texttt{Transformer}} & \multirow{2}{*}{PyTorch}    & Train                      & 154                     & 2,901 (53)                           & 547 (71) & 615K (93)   & 1,592 (72) & 7,165 (97)  \\
                                                  &                             & Inference                  & 154                     & 2,901 (53)                           & 547 (71) & 615K (93)   & 1,592 (73) & 7,165 (98)  \\ \cmidrule{2-9}
                                                  & \multirow{2}{*}{TensorFlow} & Train                      & 398                     & 2,727 (42)                           & 696 (46) & 1,043K (66) & 1,184 (70) & 8,478 (98)  \\
                                                  &                             & Inference                  & 396                     & 2,640 (40)                           & 692 (47) & 1,042K (67) & 1,103 (66) & 8,325 (97)  \\ \midrule
            \multirow{2}{*}{\texttt{LLama2}}      & vLLM                        & Inference                  & 170                     & 3,884 (48)                           & 724 (68) & 873K (93)   & 1,901 (72) & 7,690 (97)  \\ \cmidrule{2-9}
                                                  & Transformers                & Inference                  & 98                      & 2,860 (53)                           & 511 (72) & 582K (92)   & 1,592 (71) & 7,165 (97)  \\
            \bottomrule
        \end{tabular}
    }
\end{table*}

We execute the workloads listed in Table~\ref{tab:workloads} running Negativa-ML to debloat the shared libraries used by the workloads.
After debloating, we re-run the workloads using the debloated shared libraries to verify the correctness of debloating.
Specifically, we first compared outputs by the original and debloated workloads, confirming they are essentially identical.
Then, we compared final metrics such as test loss, validation loss or generated text for the original and debloated workloads.
We found that there are no differences between the output and final metrics, demonstrating that the debloating process does not affect the correctness of the workloads. We omit these comparisons due to space.

We then compare the total file sizes of the shared libraries before and after debloating to assess file size reductions.
Moreover, for CPU code, we compare both code size and the number of functions before and after debloating.
For GPU code, we also analyze code size reductions and the number of elements removed.
The results are presented in Table~\ref{tab:static-reduction}.

For the small model \texttt{MobileNetV2}, training with PyTorch involves 113 shared libraries, totaling 3762 MB.
After debloating, the total file size decreases by 55\%.
Notably, the CPU code size is reduced by 68\%, with 93\% of functions removed.
The GPU code size decreases by 75\%, with 98\% of elements removed.
The GPU code also accounts for the majority of the total file size and file size reduction.
Inference \texttt{MobileNetV2} with PyTorch shows similar reductions.

For the same \texttt{MobileNetV2} model, TensorFlow uses more shared libraries compared to PyTorch, with 253 libraries for training and 251 for inference
The total file sizes and reductions are smaller than PyTorch.
The CPU code in TensorFlow presents interesting results:
it has a larger size and a greater number of functions than PyTorch, yet the reduction is less significant, indicating that TensorFlow uses more CPU code and functions than PyTorch.
Given the fact that PyTorch can train(inference) the same model with less CPU code and functions, this suggests there may be unnecessary function calls within TensorFlow's CPU code—functions that are used but do not contribute meaningfully to the target ML workload.
The GPU code in TensorFlow also shows a significant reduction in size and elements, with over 99\% of elements removed.

For the medium model, \texttt{Transformer}, both PyTorch and TensorFlow show similar results to \texttt{MobileNetV2}, indicating that different models do not significantly affect the bloat in the ML frameworks.
For the large model, \texttt{LLama2}, the frameworks vLLM and Transformers were used for inference, and despite differences in frameworks, the reductions in file size, CPU code and GPU code remain significant.

The size of GPU code is considerably larger than that of CPU code for all ML frameworks.
GPU code also exhibits a higher reduction in both size and element count for all ML frameworks.
In particular, the element count reduction in GPU code exceeds 97\% across all workloads, underscoring that GPU code is significantly more bloated than CPU code.

\begin{tcolorbox}
    \textsc{Summary.} All ML frameworks show substantial reductions in both CPU code ($\ge{46\%}$) and GPU code ($\ge{66\%}$).
    GPU code is notably more bloated than CPU code, contributing to the majority of the bloat within the ML frameworks.
\end{tcolorbox}

\subsection{Shared Library Level Analysis}
In this section, we delve deeper into the distribution of bloat within CPU and GPU code at the shared library level.
For each shared library used in the workloads, we calculate the CPU code size reduction and function count reduction in percentage.
Similarly, for each shared library, we calculate the reduction percentage in GPU code size and element count as well.
If a shared library does not have GPU code, we exclude it from the GPU code analysis.

\begin{figure}[h]
    \begin{subfigure}[b]{0.238\textwidth}
        \centering
        \includegraphics[width=1\textwidth]{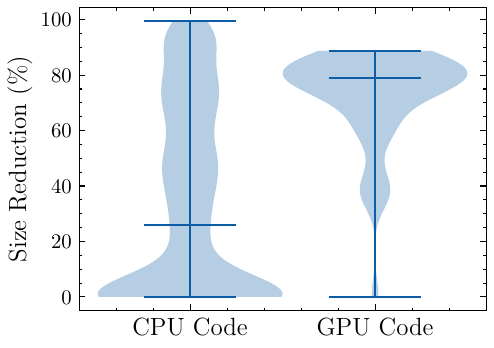}
        \caption{Distribution of CPU code size reduction and GPU code size reduction.}
        \label{fig:violin-code-size}
    \end{subfigure}
    \hfil
    \begin{subfigure}[b]{0.238\textwidth}
        \centering
        \includegraphics[width=1\textwidth]{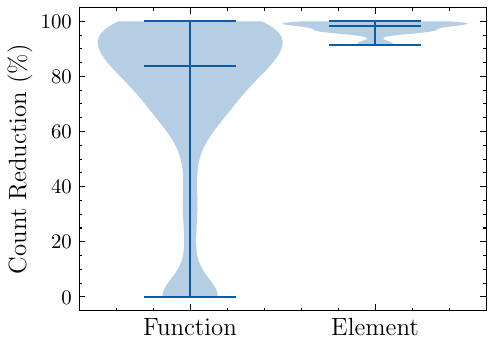}
        \caption{Distribution of CPU function count reduction and GPU element count reduction.}
        \label{fig:violin-count}
    \end{subfigure}
    \caption{Violin plots of distribution of CPU and GPU code reduction in size and function count.}
\end{figure}

Figure~\ref{fig:violin-code-size} illustrates the violin plots of distribution of size reduction in CPU and GPU code.
The distribution patterns of CPU and GPU code show distinct differences.
The median reduction for CPU code size is approximately 25\%, with many shared libraries showing a reduction from 0\% to 10\%.
In contrast, the median reduction for GPU code size is almost 80\%, significantly higher than that for CPU code, and its distribution is also more concentrated.
Figure~\ref{fig:violin-count} depicts the distribution of reductions in function and element counts.
Notably, all shared libraries exhibit an element reduction of over 80\%.
The concentrated distributions for GPU code size and element count reductions highlight that GPU code is considerably more bloated than CPU code in all shared libraries.

\begin{figure}[h]
    \centering
    \includegraphics[width=0.3\textwidth]{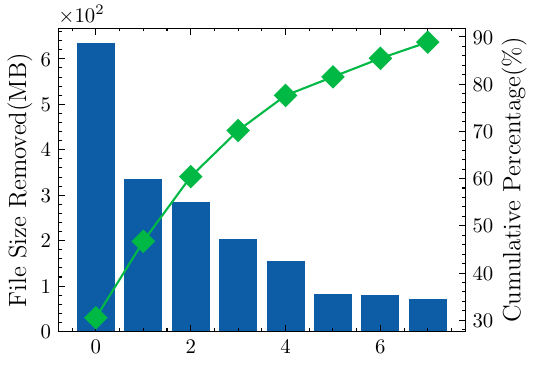}
    \caption{Pareto chart of file size reduced in the shared libraries used by the PyTorch Training \texttt{MobileNetV2} workload. The names of shared libraries are shown as indices in \textit{x-axis} for brevity.}
    \label{fig:pareto-file-size-reduction}
    \vspace{-1em}
\end{figure}

For each workload, we sorted the shared libraries by their absolute file size reductions in descending order.
We found that the top 10\% shared libraries contribute over 90\% of the total size reduction for all ML frameworks.
Reductions in CPU code size and GPU code size also follow a similar pattern.
For instance, as illustrated in Figure~\ref{fig:pareto-file-size-reduction}, the Pareto chart for the PyTorch Training \texttt{MobileNetV2} workload shows that among 113 shared libraries, the top 8 libraries account for 90\% of the total file size reduction.
This result suggests that bloat follows a Power Law distribution, with a few shared libraries containing the majority of the bloat.

\begin{tcolorbox}
    \textsc{Summary.} 50\% of shared libraries in ML frameworks have GPU code reductions more than 80\%.
    10\% of shared libraries contribute over 90\% of the total size reduction.
\end{tcolorbox}

\subsection{Function and Element Level Analysis}
\begin{table*}[ht]
    \centering
    \caption{Static size, function count and element count of the core shared libraries in ML frameworks. The table shows the original value of a metric and the reduction in percentage of the metric in parentheses. K=1000.}
    \label{tab:large-so}
    \scalebox{0.9}{
        \begin{threeparttable}
            \begin{tabular}{llllrrrrr}
                \toprule
                \multirow{2}{*}{Model}                & \multirow{2}{*}{Framework}  & \multirow{2}{*}{Operation} & \multirow{2}{*}{Lib. Name}  & \multirow{2}{*}{ \makecell{File                                                            \\ Size/MB}}              & \multicolumn{2}{c}{CPU Code} & \multicolumn{2}{c}{GPU Code}                                \\ \cmidrule(l{6pt}r{6pt}){6-7}\cmidrule(l{6pt}r{6pt}){8-9}
                                                      &                             &                            &                             &                                 & Size/MB  & \#Functions & Size/MB  & \#Elements           \\ \midrule
                \multirow{4}{*}{\texttt{MobileNetV2}} & \multirow{2}{*}{PyTorch}    & Train                      & \texttt{torch\_cuda.so}     & 841 (76)                        & 42 (91)  & 78K (93)    & 729 (82) & 2,324 (98)           \\
                                                      &                             & Inference                  & \texttt{torch\_cuda.so}     & 841 (76)                        & 42 (92)  & 78K (93)    & 729 (82) & 2,324 (98)           \\ \cmidrule{2-9}
                                                      & \multirow{2}{*}{TensorFlow} & Train                      & \texttt{tf\_cc.so}\tnote{2} & 965 (43)                        & 300 (59) & 670K (51)   & 298 (79) & 1,637 (100)\tnote{1} \\
                                                      &                             & Inference                  & \texttt{tf\_cc.so}\tnote{2} & 965 (43)                        & 300 (61) & 670K (52)   & 298 (79) & 1,637 (100)\tnote{1} \\ \midrule
                \multirow{4}{*}{\texttt{Transformer}} & \multirow{2}{*}{PyTorch}    & Train                      & \texttt{torch\_cuda.so}     & 841 (73)                        & 42 (91)  & 78K (93)    & 729 (79) & 2,324 (96)           \\
                                                      &                             & Inference                  & \texttt{torch\_cuda.so}     & 841 (76)                        & 42 (92)  & 78K (94)    & 729 (82) & 2,324 (98)           \\ \cmidrule{2-9}
                                                      & \multirow{2}{*}{TensorFlow} & Train                      & \texttt{tf\_cc.so}\tnote{2} & 965 (43)                        & 300 (59) & 670K (51)   & 298 (79) & 1,637 (100)\tnote{1} \\
                                                      &                             & Inference                  & \texttt{tf\_cc.so}\tnote{2} & 965 (41)                        & 300 (59) & 670K (52)   & 298 (73) & 1,637 (94)           \\ \midrule
                \multirow{2}{*}{\texttt{LLama2}}      & vLLM                        & Inference                  & \texttt{torch\_cuda.so}     & 861 (74)                        & 43 (91)  & 78K (93)    & 747 (80) & 2,359 (97)           \\ \cmidrule{2-9}
                                                      & Transformers                & Inference                  & \texttt{torch\_cuda.so}     & 841 (73)                        & 42 (91)  & 78K (93)    & 729 (79) & 2,324 (96)           \\ \bottomrule
            \end{tabular}
            \begin{tablenotes}
                \item[1] The reduction is 99.8\% and rounded to 100\%.
                \item[2] \texttt{tf\_cc.so} is the abbreviation of \texttt{tensorflow\_cc.so}.
            \end{tablenotes}
        \end{threeparttable}
    }
\end{table*}

\begin{table*}[]
    \centering
    \caption{Jaccard Similarity of used functions and kernels in \texttt{torch\_cuda.so} for each pair of workloads. The bottom left shows the similarity of kernels between each pair of workloads. The top right shows the similarity of functions between each pair of workloads.}
    \label{tab:similarity}
    \begin{small}
        \begin{tabular}{lllrrrrr}
            \toprule
                                                  &                          &           & \multicolumn{2}{c}{\texttt{MobileNetV2}} & \multicolumn{2}{c}{\texttt{Transformer}} & \multicolumn{1}{c}{\texttt{Llama2}}                                                        \\
                                                  &                          &           & \multicolumn{2}{c}{PyTorch}              & \multicolumn{2}{c}{PyTorch}              & \multicolumn{1}{c}{Transformers}                                                           \\
            \cmidrule(r){4-5} \cmidrule(r){6-7}  \cmidrule(r){8-8}
                                                  &                          &           & Train                                    & Inference                                & Train                               & Inference                & Inference                 \\ \midrule
            \multirow{2}{*}{\texttt{MobileNetV2}} & \multirow{2}{*}{PyTorch} & Train     & -                                        & 0.96  \cellcolor{gray!25}                & 0.89  \cellcolor{gray!25}           & 0.89 \cellcolor{gray!25} & 0.73 \cellcolor{gray!25}  \\
                                                  &                          & Inference & 0.42  \cellcolor{gray!54}                & -                                        & 0.89 \cellcolor{gray!25}            & 0.92 \cellcolor{gray!25} & 0.74  \cellcolor{gray!25} \\ \hline
            \multirow{2}{*}{\texttt{Transformer}} & \multirow{2}{*}{PyTorch} & Train     & 0.12  \cellcolor{gray!54}                & 0.06 \cellcolor{gray!54}                 & -                                   & 0.94 \cellcolor{gray!25} & 0.75  \cellcolor{gray!25} \\
                                                  &                          & Inference & 0.06    \cellcolor{gray!54}              & 0.13 \cellcolor{gray!54}                 & 0.24  \cellcolor{gray!54}           & -                        & 0.77  \cellcolor{gray!25} \\ \hline
            \texttt{LLama2}                       & Transformers             & Inference & 0.07   \cellcolor{gray!54}               & 0.08  \cellcolor{gray!54}                & 0.07 \cellcolor{gray!54}            & 0.08 \cellcolor{gray!54} & -                         \\ \bottomrule
        \end{tabular}
    \end{small}
\end{table*}

\begin{table*}[]
    \centering
    \caption{Average runtime performance using original shared libraries and reductions using debloated shared libraries. The numbers in parentheses are the percentage of reduction.
        The standard deviation of all metrics for each workload is less than 2\% and is not shown.}
    \label{tab:runtime-performance}
    \scalebox{0.97}{

        \begin{tabular}{lllrrr}
            \toprule
            Model                                                       & Framework                   & Operation       & Peak CPU Memory/MB & Peak GPU Memory/MB & Execution Time/s \\ \midrule
            \multirow{4}{*}{\texttt{MobileNetV2}}                       & \multirow{2}{*}{PyTorch}    & Train           & 5,487 (64.2)       & 1,539  (48.1)      & 179  (2.3)       \\
                                                                        &                             & Inference       & 4,943 (74.6)       & 972  (69.6)        & 8  (44.6)        \\ \cmidrule{2-6}
                                                                        & \multirow{2}{*}{TensorFlow} & Train           & 6,009 (48.7)       & 14,395  (2.8)      & 53  (5.5)        \\
                                                                        &                             & Inference       & 4,850 (60.0)       & 14,323  (2.8)      & 12  (24.2)       \\ \midrule
            \multirow{4}{*}{\texttt{Transformer}}                       & \multirow{2}{*}{PyTorch}    & Train           & 4,151 (56.0)       & 9,381  (5.0)       & 200  (1.1)       \\
                                                                        &                             & Inference       & 4,054 (65.8)       & 1,349  (42.3)      & 13  (20.3)       \\ \cmidrule{2-6}
                                                                        & \multirow{2}{*}{TensorFlow} & Train           & 15,652 (13.6)      & 14,149  (1.6)      & 4,779  (0.0)     \\
                                                                        &                             & Inference       & 4,217 (36.5)       & 14,069  (1.3)      & 69  (1.8)        \\ \midrule
            \multirow{2}{*}{\texttt{Llama2}}                            & vLLM                        & Inference       & 12,527 (11.8)      & 14,679  (2.1)      & 43  (12.7)       \\ \cmidrule{2-6}
                                                                        & Transformers                & Inference       & 12,065 (10.4)      & 13,793  (3.2)      & 21  (8.9)        \\ \midrule
            \multicolumn{3}{c}{Average Absolute Reduction$\pm{\sigma}$} & 2501$\pm{825}$              & 443  $\pm{171}$ & 2.6$\pm1.6$                                                \\ \bottomrule
        \end{tabular}
    }
\end{table*}

To deepen the analysis, Table~\ref{tab:large-so} presents reductions for the largest shared library used in each workload.
All the four ML frameworks while running the different models use one of two shared libraries, \texttt{torch\_cuda.so} and \texttt{tensorflow\_cc.so}.
These two libraries, which provide the core functionalities of the ML frameworks, also exhibit significant reductions in file size, GPU code, and CPU code.
For the \texttt{torch\_cuda.so}, the reduction in file size, CPU code size, and GPU code size are 76\%, 91\% and 82\%, respectively.
The CPU code in \texttt{tensorflow\_cc.so} also presents the interesting results as we have discussed in \S\ref{sec:overview}:
it has a much larger CPU code size and a greater number of functions than \texttt{torch\_cuda.so}, yet the reduction is smaller.
The reductions in GPU code for \texttt{tensorflow\_cc.so} is as significant as \texttt{torch\_cuda.so}.

In the table, \texttt{torch\_cuda.so} is used by three ML frameworks, PyTorch, vLLM, and Transformers, involving 6 workloads.
We collect the functions used by each workload in \texttt{torch\_cuda.so} to compare their similarity.
Since vLLM uses a different version of \texttt{torch\_cuda.so}, we exclude it from the analysis.
Therefore, in total we collect five sets of functions used by five different workloads.
For each pair of the function sets, we calculate their Jaccard Similarity according to the following formula:
\begin{equation}
    J(A, B) = \frac{|A \cap B|}{|A \cup B|}
\end{equation}
where $A$ and $B$ are two sets of functions used by two workloads.
The Jaccard Similarity is 1 if the two sets are the same, and 0 if the two sets are disjoint.
The Jaccard Similarity is also calculated for the kernels used by the five workloads.
The results are shown in Table~\ref{tab:similarity}.
We also did the same analysis for \texttt{tensorflow\_cc.so}, and the results are similar and shown in the appendix.

The bottom left part of Table~\ref{tab:similarity} shows the similarity of kernels between each pair of workloads.
The similarity is quite low, indicating that the kernels used by different workloads are quite different.
However, the similarity of functions between each pair of workloads is very high, as shown in the upper right part of Table~\ref{tab:similarity}.
All pairs have a similarity above 0.7, indicating different workloads used many common functions, even if the workloads run with different models and frameworks.
Given the fact that less than 10\% of functions are actually used in the \texttt{torch\_cuda.so} library in each of the five workloads, this suggests that a small subset of functions is sufficient to run various workloads.

\begin{figure}[]
    \centering
    \includegraphics[width=0.42\textwidth]{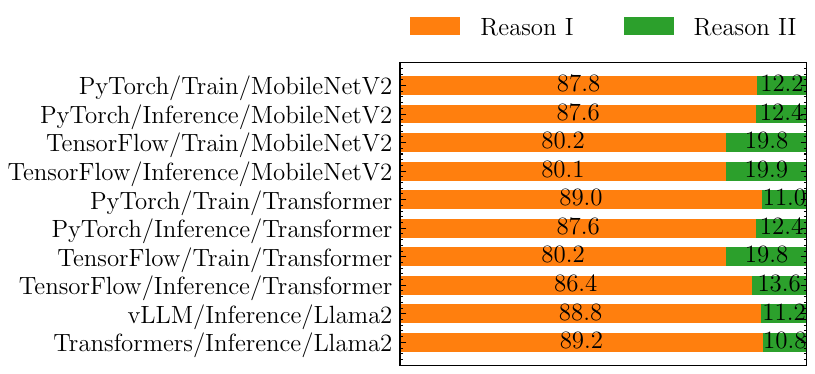}
    \caption{Reasons of removed elements in GPU code.}
    \label{fig:element-reason}
\end{figure}

We also examine the reasons for removed elements in GPU code.
Negativa-ML removes an element due to one of the following two reasons:
\textbf{Reason I}: The element does not match the GPU architecture; \textbf{Reason II}:  The element matches the GPU architecture, but it does not have any used kernels.
We identify the reason for each removed element.

Figure~\ref{fig:element-reason} shows that for all workloads, over 80\% of removed elements are due to Reason I, i.e., the element does not match the GPU architecture.
The cause of these unmatched elements is that the GPU code in these shared libraries are built to support various GPU architectures, leading to numerous unnecessary elements for workloads on a specific GPU.
For example, in our experiment, we observed a single shared library in the PyTorch framework contained elements for 6 different GPU architectures.
If the library is deployed on a specific GPU architecture, the elements for the five other architectures are unnecessary.
This hints at a new reason for bloat: software bloat can stem from hardware.

\begin{tcolorbox}
    \textsc{Summary.} Different workloads used a lot of common functions in the core shared libraries. Most elements removed in GPU code are due to the mismatch of GPU architecture.
\end{tcolorbox}

\subsection{Runtime Performance Analysis}

In this section, we evaluate runtime performance improvements after debloating.
Initially, each workload is executed ten times using the original shared libraries.
Then, based on the previous finding that a few shared libraries contribute the majority of bloat, we replace the top 8 largest shared libraries with their debloated versions and re-ran the workloads ten times.
We then compare the average runtime performance between the two runs.
Table~\ref{tab:runtime-performance} summarizes the average runtime performance improvements achieved with the debloated shared libraries.

Running workloads with debloated libraries reduces memory usage (both CPU and GPU) and execution time.
The best improvement is observed in the PyTorch Inference \texttt{MobileNetV2} workload, with reductions of 74.6\% in peak CPU memory, 69.6\% in peak GPU memory, and 44.6\% in execution time.
Inference workloads generally show better improvement than training workloads.
Across all workloads, average absolute reductions are 2501 MB for peak CPU memory, 443 MB for peak GPU memory, and 2.6 seconds for execution time.
Notably, the absolute execution time reduction remains relatively constant, regardless of the actual execution duration.
This improvement stems from reduced code size, which decreases the time required to load the code into memory.
This execution time improvement is especially impactful for tasks sensitive to cold start latency, such as serverless ML applications.

\begin{tcolorbox}
    \textsc{Summary.} Debloating shared libraries significantly reduces both CPU and GPU memory usage.
    The time required to load the code into memory is also reduced, leading to shorter execution times.
\end{tcolorbox}

\subsection{Evaluation on Different GPUs}

\begin{table*}[]
    \centering
    \caption{Total file size, CPU code, GPU code and their reductions of all shared libraries in vLLM/Transformers inference \texttt{Llama2} workload on 1 H100 GPU under eager-loading and lazy-loading modes. The table shows the original value of a metric and the reduction in percentage of the metric in parentheses. K=1000.}
    \label{tab:h100-static-reduction}
    \scalebox{0.96}{
        \begin{tabular}{llrrrrrr}
            \toprule
            \multirow{2}{*}{Framework}    & \multirow{2}{*}{Mode} & \multirow{2}{*}{\#Lib.} & \multirow{2}{*}{\makecell{Total File                                                    \\ Size/MB}}                      & \multicolumn{2}{c}{CPU Code} & \multicolumn{2}{c}{GPU Code}                        \\ \cmidrule(l{6pt}r{6pt}){5-6}\cmidrule(l{6pt}r{6pt}){7-8}
                                          &                       &                         &                                      & Size/MB  & \#Functions & Size/MB    & \#Elements \\ \hline
            \multirow{2}{*}{vLLM}         & Eager                 & 170                     & 4,068 (47)                           & 726 (66) & 872K (93)   & 2,104 (68) & 7,692 (89) \\ \cmidrule{2-8}
                                          & Lazy                  & 170                     & 4,068 (47)                           & 726 (66) & 872K (93)   & 2,104 (68) & 7,692 (89) \\ \midrule
            \multirow{2}{*}{Transformers} & Eager                 & 93                      & 2,868 (51)                           & 513 (72) & 584K (92)   & 1,592 (69) & 7,165 (88) \\ \cmidrule{2-8}
                                          & Lazy                  & 93                      & 2,868 (51)                           & 513 (72) & 584K (92)   & 1,592 (69) & 7,165 (88) \\
            \bottomrule
        \end{tabular}
    }
\end{table*}

\begin{table*}[]
    \centering
    \caption{Average runtime performance using original shared libraries and reductions using debloated shared libraries in vLLM/Transformers inference \texttt{Llama2} workload on 1 H100 GPU under eager-loading and lazy-loading modes. The numbers in
        parentheses are the percentage of reduction. The standard deviation of all metrics for each workload is less than 2\% and is not shown.}
    \label{tab:h100-runtime-reduction}
    \scalebox{0.96}{
        \begin{tabular}{llrrr}
            \toprule
            Framework    & Mode  & Peak CPU Memory/MB & Peak GPU Memory/MB & Execution Time/s \\ \midrule
            vLLM         & Eager & 13,333 (18.0)      & 91,946  (0.7)      & 44  (13.9)       \\ \cmidrule{2-5}
                         & Lazy  & 10,313 (0.3)       & 91,061  (0.0)      & 36  (8.3)        \\ \midrule
            Transformers & Eager & 12,345 (12.2)      & 15,002  (2.4)      & 23  (32.0)       \\ \cmidrule{2-5}
                         & Lazy  & 10,049 (0.2)       & 13,764  (0.0)      & 17  (20.3)       \\
            \bottomrule
        \end{tabular}
    }

\end{table*}

In this section, we present debloating results of the two LLM inference frameworks, vLLM and Transformers, using different GPUs.
We first evaluate the vLLM/Transformers inference workload using \texttt{Llama2} on a single H100 GPU.
The evaluation is conducted under both eager-loading and lazy-loading modes.
In eager-loading, all kernels are loaded into memory at application startup, whereas lazy-loading loads kernels only when needed, reducing memory footprint.
Table~\ref{tab:h100-static-reduction} presents the size reductions achieved. For the same framework, the reductions in file size, CPU code size, function count, and GPU code size are similar between the two loading modes.
The runtime performance improvements are shown in Table~\ref{tab:h100-runtime-reduction}.
Under eager-loading, debloating results in greater CPU memory reduction compared to lazy-loading, while GPU memory reduction remains similar across both modes. Both loading modes exhibit reductions in execution time.
Comparing these results with those obtained using the T4 GPU (Table~\ref{tab:static-reduction} and Table~\ref{tab:runtime-performance}), we observe consistent reductions across different GPUs.
This demonstrates that Negativa-ML effectively debloats shared libraries across various GPU architectures.

Next, we present debloating results for additional ML workloads using vLLM and Transformers.
We select the top 9 LLMs from the Hugging Face Open LLM Leaderboard~\cite{open-llm-leaderboard-v1} and deploy them on both frameworks using distributed inference with 8×A100 40GB GPUs.
This setup aims to evaluate whether debloating works with distributed inference under a different hardware setup.
Results are detailed in Table~\ref{tab:general_results} in the Appendix.
The debloating results of distributed inference with 8 GPUs are still as significant as those of single-GPU inference.
For inference with 8 GPUs, vLLM uses slightly fewer shared libraries than the results of a single GPU, while Transformers uses nearly the same number of libraries.
The reductions in file size, CPU code size, function count, and GPU code size align closely with the results using a single GPU.
However, the element count reduction in GPU code is lower than that of single-GPU inference, suggesting that distributed inference utilizes more kernels in GPU code.
Additionally, the reduction metrics are similar across different models, once again indicating that different models do not significantly affect the bloat in the ML frameworks.

\begin{tcolorbox}
    \textsc{Summary.} Negativa-ML effectively debloats ML shared libraries across different GPUs and distributed inference setups.
\end{tcolorbox}

\subsection{Performance of Negativa-ML}
In this section, we evaluate the performance of Negativa-ML in terms of the time taken to debloat shared libraries.
We first evaluate the end-to-end time, i.e., the time taken from running the target ML workload until obtaining the debloated shared libraries.
Then, we specifically evaluate the overhead introduced to the target ML workload by the kernel detector.

Table~\ref{tab:end-to-end-negativa} shows the end-to-end time taken by Negativa-ML to debloat shared libraries for each workload.
The time varies depending on three factors: the original execution time of the workload, the number of shared libraries involved, and the number of functions (kernels) utilized by the workload.
We note that the times taken are to debloat all the libraries in the workload.
If only debloating a single library, the time needed will be significantly shorter.
Moreover, the debloating process is a one-time overhead that can occur in, e.g., a preprocessing stage prior to deployment.

\begin{table}[htbp]
    \centering
    \caption{End-to-end time taken by Negativa-ML to debloat shared libraries for each workload.}
    \label{tab:end-to-end-negativa}
    \scalebox{0.87}{
        \begin{tabular}{lllrr}
            \toprule
            Model                                 & Framework                   & Operation & \#Lib. & Time/s \\ \midrule
            \multirow{4}{*}{\texttt{MobileNetV2}} & \multirow{2}{*}{PyTorch}    & Train     & 113    & 651    \\
                                                  &                             & Inference & 111    & 383    \\ \cmidrule{2-5}
                                                  & \multirow{2}{*}{TensorFlow} & Train     & 253    & 659    \\
                                                  &                             & Inference & 251    & 585    \\ \midrule
            \multirow{4}{*}{\texttt{Transformer}} & \multirow{2}{*}{PyTorch}    & Train     & 154    & 1,247  \\
                                                  &                             & Inference & 154    & 344    \\ \cmidrule{2-5}
                                                  & \multirow{2}{*}{TensorFlow} & Train     & 398    & 18,420 \\
                                                  &                             & Inference & 396    & 639    \\ \midrule
            \multirow{2}{*}{\texttt{Llama2}}      & vLLM                        & Inference & 170    & 713    \\ \cmidrule{2-5}
                                                  & Transformers                & Inference & 98     & 362    \\ \bottomrule
        \end{tabular}
    }
\end{table}

To assess the performance overhead introduced by the kernel detector on the target ML workload, we first run the PyTorch Training \texttt{MobileNetV2} workload 10 times and recorded the average execution time.
Next, we ran the same workload, this time with tracing enabled by the kernel detector and by NSys respectively, each for 10 runs.
We then compare the average execution times across the three setups.
The average execution time for the original workload was 180 seconds.
With the kernel detector enabled, the execution time increased to 253 seconds, a 41\% overhead.
In contrast, tracing with NSys increases the execution time to 407 seconds, introducing a 126\% overhead.
The kernel detector imposes significantly lower overhead than NSys, making it a more practical choice for detecting used kernels, especially for long-running workloads like ML training. 

\begin{tcolorbox}
    \textsc{Summary.} Debloating is a one-time overhead that can be performed in a preprocessing stage prior to deployment. The kernel detector introduces a 41\% overhead for the first run, significantly lower than traditional tracing tools like NSys which introduce a 126\% overhead.
\end{tcolorbox}
\section{Discussion}
As Sculley et al.~\cite{sculley2015hidden} highlight, ML systems have hidden technical debt due to a set of ML-specific issues.
Our work emphasizes this technical debt by exposing the hidden bloat in ML frameworks.
Software bloat has been a long-standing issue in the software industry.
While existing debloating research has focused on traditional software, bloat in ML systems is not well understood.
The uniqueness of ML systems is that they contain both CPU and GPU code, leading to significant bloat.
Our evaluations over four ML frameworks across ten workloads with around 300 shared libraries show significant bloat in these frameworks, with up to 72\% size reduction in CPU code and 75\% size reduction in GPU code.
Furthermore, unlike traditional software, ML frameworks experience substantial bloat in GPU code.
All this bloat leads to increased storage needs, higher memory usage, and longer execution times.

In our evaluations, only a small subset of code is consistently utilized. This suggests that code unused by one workload is likely unnecessary for others as well.
Besides, existing research focuses solely on unused code, overlooking ``used bloat” — code executed by a workload but not contributing meaningfully to the performance or functionality.
For instance, when training a model with a specific optimizer, the optimizer may initialize a context with numerous non-essential function calls.
Compared with PyTorch, the larger CPU code size but smaller reduction in TensorFlow may indicate the presence of ``used bloat''.
Such ``used bloat'' is particularly harmful, as it is executed and thus consumes scarce memory, CPU, and GPU resources.
Moreover, it is more difficult to detect as it is actually executed.
Future research could focus on identifying and eliminating this ``used bloat''.

Shared libraries in ML frameworks are significantly larger than traditional shared libraries~\cite{zhang2024machine}, mainly due to GPU code as shown in this work.
As storage and network bandwidth are critical bottlenecks in edge data centers~\cite{richins2021ai}, the file size reduction achieved by debloating can help alleviate these bottlenecks.


\section{Conclusion}
We propose a novel approach to debloat ML frameworks by removing unnecessary code in both CPU and GPU code within shared libraries.
Our approach first detects kernel names used in GPU code by running ML workloads, such as training or inference a model.
We then locate the used kernels within the shared libraries and remove the unused code.
We implement our approach based on an existing debloating tool, and evaluate it on four ML frameworks across ten workloads over 300 shared libraries.
Our evaluation shows that ML frameworks are highly bloated, with up to 55\% size reduction in shared libraries in these frameworks.
We show that up to 75\% of GPU code and 99\% of GPU elements being unnecessary for target ML workloads;
Up to 72\% of CPU code and 93\% of CPU functions are unnecessary for target ML workloads.
This bloat not only increases storage overhead but also degrades runtime performance.
Debloating these frameworks reduces the peak memory usage, peak GPU memory usage, and startup time by up to 74.6\%, 69.6\%, and 44.6\%, respectively.

 \noindent\textbf{Acknowledgment} \small This work is funded by a WASP PhD student grant and an SSF grant (grant \#FFL21-0091).
\clearpage

\nocite{langley00}

\bibliography{example_paper}

\begin{thebibliography}{47}
\providecommand{\natexlab}[1]{#1}
\providecommand{\url}[1]{\texttt{#1}}
\expandafter\ifx\csname urlstyle\endcsname\relax
  \providecommand{\doi}[1]{doi: #1}\else
  \providecommand{\doi}{doi: \begingroup \urlstyle{rm}\Url}\fi

\bibitem[hug()]{huggingfaceTextGeneration}
{T}ext {G}eneration {I}nference --- huggingface.co.
\newblock \url{https://huggingface.co/docs/text-generation-inference/index}.
\newblock [Accessed 28-10-2024].

\bibitem[nvi({\natexlab{a}})]{nvidiaCUDABinary}
{C}{U}{D}{A} {B}inary {U}tilities --- docs.nvidia.com.
\newblock \url{https://docs.nvidia.com/cuda/cuda-binary-utilities/index.html}, {\natexlab{a}}.
\newblock [Accessed 28-10-2024].

\bibitem[nvi({\natexlab{b}})]{nvidiaCUDADeep}
{C}{U}{D}{A} {D}eep {N}eural {N}etwork --- developer.nvidia.com.
\newblock \url{https://developer.nvidia.com/cudnn}, {\natexlab{b}}.
\newblock [Accessed 28-10-2024].

\bibitem[nvi({\natexlab{c}})]{nvidiaCuBLAS}
cu{B}{L}{A}{S} --- developer.nvidia.com.
\newblock \url{https://developer.nvidia.com/cublas}, {\natexlab{c}}.
\newblock [Accessed 28-10-2024].

\bibitem[nvi({\natexlab{d}})]{nvidiaCupti}
{N}{V}{I}{D}{I}{A} {C}{U}{D}{A} {P}rofiling {T}ools {I}nterface ({C}{U}{P}{T}{I}) - {C}{U}{D}{A} {T}oolkit --- developer.nvidia.com.
\newblock \url{https://developer.nvidia.com/cupti}, {\natexlab{d}}.
\newblock [Accessed 28-10-2024].

\bibitem[nvi({\natexlab{e}})]{nvidiaNVIDIACUDA}
{N}{V}{I}{D}{I}{A} {C}{U}{D}{A} {C}ompiler {D}river --- docs.nvidia.com.
\newblock \url{https://docs.nvidia.com/cuda/cuda-compiler-driver-nvcc/index.html#using-separate-compilation-in-cuda}, {\natexlab{e}}.
\newblock [Accessed 27-10-2024].

\bibitem[nvi({\natexlab{f}})]{nvidiaNVIDIANsys}
{N}{V}{I}{D}{I}{A} {N}sight {S}ystems --- developer.nvidia.com.
\newblock \url{https://developer.nvidia.com/nsight-systems}, {\natexlab{f}}.
\newblock [Accessed 28-10-2024].

\bibitem[ora()]{oracleFileFormat}
{F}ile {F}ormat ({L}inker and {L}ibraries {G}uide) --- docs.oracle.com.
\newblock \url{https://docs.oracle.com/cd/E19683-01/816-1386/6m7qcoblj/index.html}.
\newblock [Accessed 28-10-2024].

\bibitem[Abadi et~al.(2016)Abadi, Barham, Chen, Chen, Davis, Dean, Devin, Ghemawat, Irving, Isard, et~al.]{abadi2016tensorflow}
Abadi, M., Barham, P., Chen, J., Chen, Z., Davis, A., Dean, J., Devin, M., Ghemawat, S., Irving, G., Isard, M., et~al.
\newblock $\{$TensorFlow$\}$: a system for $\{$Large-Scale$\}$ machine learning.
\newblock In \emph{12th USENIX symposium on operating systems design and implementation (OSDI 16)}, pp.\  265--283, 2016.

\bibitem[Agadakos et~al.(2019)Agadakos, Jin, Williams-King, Kemerlis, and Portokalidis]{agadakos2019nibbler}
Agadakos, I., Jin, D., Williams-King, D., Kemerlis, V.~P., and Portokalidis, G.
\newblock Nibbler: debloating binary shared libraries.
\newblock In \emph{Proceedings of the 35th Annual Computer Security Applications Conference}, pp.\  70--83, 2019.

\bibitem[Ahmad et~al.(2021)Ahmad, Noor, Sharif, Hameed, Asif, Anwar, Gehani, Zaffar, and Siddiqui]{ahmad2021trimmer}
Ahmad, A.~A., Noor, A.~R., Sharif, H., Hameed, U., Asif, S., Anwar, M., Gehani, A., Zaffar, F., and Siddiqui, J.~H.
\newblock Trimmer: an automated system for configuration-based software debloating.
\newblock \emph{IEEE Transactions on Software Engineering}, 48\penalty0 (9):\penalty0 3485--3505, 2021.

\bibitem[Azad et~al.(2019)Azad, Laperdrix, and Nikiforakis]{azad2019less}
Azad, B.~A., Laperdrix, P., and Nikiforakis, N.
\newblock Less is more: Quantifying the security benefits of debloating web applications.
\newblock In \emph{28th USENIX Security Symposium (USENIX Security 19)}, pp.\  1697--1714, 2019.

\bibitem[Beeching et~al.(2023)Beeching, Fourrier, Habib, Han, Lambert, Rajani, Sanseviero, Tunstall, and Wolf]{open-llm-leaderboard-v1}
Beeching, E., Fourrier, C., Habib, N., Han, S., Lambert, N., Rajani, N., Sanseviero, O., Tunstall, L., and Wolf, T.
\newblock Open llm leaderboard (2023-2024).
\newblock \url{https://huggingface.co/spaces/open-llm-leaderboard/open_llm_leaderboard}, 2023.

\bibitem[Bhardwaj et~al.(2017)Bhardwaj, Nambiar, and Dutta]{bhardwaj2017study}
Bhardwaj, R., Nambiar, A.~R., and Dutta, D.
\newblock A study of machine learning in healthcare.
\newblock In \emph{2017 IEEE 41st annual computer software and applications conference (COMPSAC)}, volume~2, pp.\  236--241. IEEE, 2017.

\bibitem[Biswas et~al.(2021)Biswas, Burow, and Payer]{biswas2021ancile}
Biswas, P., Burow, N., and Payer, M.
\newblock Code specialization through dynamic feature observation.
\newblock In \emph{Proceedings of the Eleventh ACM Conference on Data and Application Security and Privacy}, pp.\  257--268, 2021.

\bibitem[Bojar et~al.(2014)Bojar, Buck, Federmann, Haddow, Koehn, Leveling, Monz, Pecina, Post, Saint-Amand, Soricut, Specia, and Tamchyna]{bojar-EtAl:2014:W14-33}
Bojar, O., Buck, C., Federmann, C., Haddow, B., Koehn, P., Leveling, J., Monz, C., Pecina, P., Post, M., Saint-Amand, H., Soricut, R., Specia, L., and Tamchyna, A.~s.
\newblock Findings of the 2014 workshop on statistical machine translation.
\newblock In \emph{Proceedings of the Ninth Workshop on Statistical Machine Translation}, pp.\  12--58, Baltimore, Maryland, USA, June 2014. Association for Computational Linguistics.
\newblock URL \url{http://www.aclweb.org/anthology/W/W14/W14-3302}.

\bibitem[Brown \& Pande(2019)Brown and Pande]{brown2019carve}
Brown, M.~D. and Pande, S.
\newblock Carve: Practical security-focused software debloating using simple feature set mappings.
\newblock In \emph{Proceedings of the 3rd ACM Workshop on Forming an Ecosystem Around Software Transformation}, pp.\  1--7, 2019.

\bibitem[Brown et~al.(2024)Brown, Meily, Fairservice, Sood, Dorn, Kilmer, and Eytchison]{brown2024broad}
Brown, M.~D., Meily, A., Fairservice, B., Sood, A., Dorn, J., Kilmer, E., and Eytchison, R.
\newblock A broad comparative evaluation of software debloating tools.
\newblock pp.\  3927--3943, 2024.

\bibitem[Elliott et~al.(2016)Elliott, Frank, Sima'an, and Specia]{W16-3210}
Elliott, D., Frank, S., Sima'an, K., and Specia, L.
\newblock Multi30k: Multilingual english-german image descriptions.
\newblock In \emph{Proceedings of the 5th Workshop on Vision and Language}, pp.\  70--74. Association for Computational Linguistics, 2016.
\newblock \doi{10.18653/v1/W16-3210}.
\newblock URL \url{http://www.aclweb.org/anthology/W16-3210}.

\bibitem[Fernandez et~al.(2023)Fernandez, Kahn, Na, Bisk, and Strubell]{fernandez2023framework}
Fernandez, J., Kahn, J., Na, C., Bisk, Y., and Strubell, E.
\newblock The framework tax: Disparities between inference efficiency in nlp research and deployment.
\newblock In \emph{Proceedings of the 2023 Conference on Empirical Methods in Natural Language Processing}, pp.\  1588--1600, Singapore, December 2023. Association for Computational Linguistics.

\bibitem[Jiang et~al.(2020)Jiang, Wang, Chen, Wu, Wang, Zou, Yang, Cui, Cai, Yu, et~al.]{jiang2020mnn}
Jiang, X., Wang, H., Chen, Y., Wu, Z., Wang, L., Zou, B., Yang, Y., Cui, Z., Cai, Y., Yu, T., et~al.
\newblock Mnn: A universal and efficient inference engine.
\newblock \emph{Proceedings of Machine Learning and Systems}, 2:\penalty0 1--13, 2020.

\bibitem[Ko et~al.(2022)Ko, Lee, Park, and Choi]{ko2022survey}
Ko, H., Lee, S., Park, Y., and Choi, A.
\newblock A survey of recommendation systems: recommendation models, techniques, and application fields.
\newblock \emph{Electronics}, 11\penalty0 (1):\penalty0 141, 2022.

\bibitem[Krizhevsky et~al.(2009)Krizhevsky, Hinton, et~al.]{krizhevsky2009learning}
Krizhevsky, A., Hinton, G., et~al.
\newblock Learning multiple layers of features from tiny images.
\newblock 2009.

\bibitem[Kuo et~al.(2020)Kuo, Chen, Mohan, and Xu]{kuo2020set}
Kuo, H.-C., Chen, J., Mohan, S., and Xu, T.
\newblock Set the configuration for the heart of the os: On the practicality of operating system kernel debloating.
\newblock \emph{Proceedings of the ACM on Measurement and Analysis of Computing Systems}, 4\penalty0 (1):\penalty0 1--27, 2020.

\bibitem[Kwon et~al.(2023)Kwon, Li, Zhuang, Sheng, Zheng, Yu, Gonzalez, Zhang, and Stoica]{kwon2023efficient}
Kwon, W., Li, Z., Zhuang, S., Sheng, Y., Zheng, L., Yu, C.~H., Gonzalez, J.~E., Zhang, H., and Stoica, I.
\newblock Efficient memory management for large language model serving with pagedattention.
\newblock In \emph{Proceedings of the ACM SIGOPS 29th Symposium on Operating Systems Principles}, 2023.

\bibitem[{LMDeploy Contributors}(2023)]{2023lmdeploy}
{LMDeploy Contributors}.
\newblock Lmdeploy: A toolkit for compressing, deploying, and serving llm.
\newblock \url{https://github.com/InternLM/lmdeploy}, 2023.

\bibitem[Navas \& Gehani(2023)Navas and Gehani]{navas2023occamv2}
Navas, J.~A. and Gehani, A.
\newblock Occam-v2: combining static and dynamic analysis for effective and efficient whole-program specialization.
\newblock \emph{Communications of the ACM}, 66\penalty0 (4):\penalty0 40--47, 2023.

\bibitem[{OpenAI Team}(2024)]{openai2024gpt4technicalreport}
{OpenAI Team}.
\newblock Gpt-4 technical report, 2024.
\newblock URL \url{https://arxiv.org/abs/2303.08774}.

\bibitem[Parekh et~al.(2022)Parekh, Poddar, Rajpurkar, Chahal, Kumar, Joshi, and Cho]{parekh2022review}
Parekh, D., Poddar, N., Rajpurkar, A., Chahal, M., Kumar, N., Joshi, G.~P., and Cho, W.
\newblock A review on autonomous vehicles: Progress, methods and challenges.
\newblock \emph{Electronics}, 11\penalty0 (14):\penalty0 2162, 2022.

\bibitem[Paszke et~al.(2019)Paszke, Gross, Massa, Lerer, Bradbury, Chanan, Killeen, Lin, Gimelshein, Antiga, et~al.]{paszke2019pytorch}
Paszke, A., Gross, S., Massa, F., Lerer, A., Bradbury, J., Chanan, G., Killeen, T., Lin, Z., Gimelshein, N., Antiga, L., et~al.
\newblock Pytorch: An imperative style, high-performance deep learning library.
\newblock \emph{Advances in neural information processing systems}, 32, 2019.

\bibitem[Qian et~al.(2019)Qian, Hu, Alharthi, Chung, Kim, and Lee]{qian2019razor}
Qian, C., Hu, H., Alharthi, M., Chung, P.~H., Kim, T., and Lee, W.
\newblock $\{$RAZOR$\}$: A framework for post-deployment software debloating.
\newblock In \emph{28th USENIX security symposium (USENIX Security 19)}, pp.\  1733--1750, 2019.

\bibitem[Quach et~al.(2017)Quach, Erinfolami, Demicco, and Prakash]{quach2017multi}
Quach, A., Erinfolami, R., Demicco, D., and Prakash, A.
\newblock A multi-os cross-layer study of bloating in user programs, kernel and managed execution environments.
\newblock In \emph{Proceedings of the 2017 Workshop on Forming an Ecosystem Around Software Transformation}, pp.\  65--70, 2017.

\bibitem[Quach et~al.(2018)Quach, Prakash, and Yan]{quach2018dpiecewise}
Quach, A., Prakash, A., and Yan, L.
\newblock Debloating software through $\{$Piece-Wise$\}$ compilation and loading.
\newblock In \emph{27th USENIX security symposium (USENIX Security 18)}, pp.\  869--886, 2018.

\bibitem[Rastogi et~al.(2017)Rastogi, Davidson, De~Carli, Jha, and McDaniel]{rastogi2017cimplifier}
Rastogi, V., Davidson, D., De~Carli, L., Jha, S., and McDaniel, P.
\newblock Cimplifier: automatically debloating containers.
\newblock In \emph{Proceedings of the 2017 11th Joint Meeting on Foundations of Software Engineering}, pp.\  476--486, 2017.

\bibitem[Richins et~al.(2021)Richins, Doshi, Blackmore, Nair, Pathapati, Patel, Daguman, Dobrijalowski, Illikkal, Long, et~al.]{richins2021ai}
Richins, D., Doshi, D., Blackmore, M., Nair, A.~T., Pathapati, N., Patel, A., Daguman, B., Dobrijalowski, D., Illikkal, R., Long, K., et~al.
\newblock Ai tax: The hidden cost of ai data center applications.
\newblock \emph{ACM Transactions on Computer Systems (TOCS)}, 37\penalty0 (1-4):\penalty0 1--32, 2021.

\bibitem[Sandler et~al.(2018)Sandler, Howard, Zhu, Zhmoginov, and Chen]{sandler2018mobilenetv2}
Sandler, M., Howard, A., Zhu, M., Zhmoginov, A., and Chen, L.-C.
\newblock Mobilenetv2: Inverted residuals and linear bottlenecks.
\newblock In \emph{Proceedings of the IEEE conference on computer vision and pattern recognition}, pp.\  4510--4520, 2018.

\bibitem[Sculley et~al.(2015)Sculley, Holt, Golovin, Davydov, Phillips, Ebner, Chaudhary, Young, Crespo, and Dennison]{sculley2015hidden}
Sculley, D., Holt, G., Golovin, D., Davydov, E., Phillips, T., Ebner, D., Chaudhary, V., Young, M., Crespo, J.-F., and Dennison, D.
\newblock Hidden technical debt in machine learning systems.
\newblock \emph{Advances in neural information processing systems}, 28, 2015.

\bibitem[Sinha et~al.(2024)Sinha, Menon, and Sagar]{sinha2024llmops}
Sinha, M., Menon, S., and Sagar, R.
\newblock Llmops: Definitions, framework and best practices.
\newblock In \emph{2024 International Conference on Electrical, Computer and Energy Technologies (ICECET}, pp.\  1--6. IEEE, 2024.

\bibitem[Touvron et~al.(2023{\natexlab{a}})Touvron, Lavril, Izacard, Martinet, Lachaux, Lacroix, Rozi{\`e}re, Goyal, Hambro, Azhar, et~al.]{touvron2023llama}
Touvron, H., Lavril, T., Izacard, G., Martinet, X., Lachaux, M.-A., Lacroix, T., Rozi{\`e}re, B., Goyal, N., Hambro, E., Azhar, F., et~al.
\newblock Llama: Open and efficient foundation language models.
\newblock \emph{arXiv preprint arXiv:2302.13971}, 2023{\natexlab{a}}.

\bibitem[Touvron et~al.(2023{\natexlab{b}})Touvron, Lavril, Izacard, Martinet, Lachaux, Lacroix, Rozière, Goyal, Hambro, Azhar, Rodriguez, Joulin, Grave, and Lample]{touvron2023llamaopenefficientfoundation}
Touvron, H., Lavril, T., Izacard, G., Martinet, X., Lachaux, M.-A., Lacroix, T., Rozière, B., Goyal, N., Hambro, E., Azhar, F., Rodriguez, A., Joulin, A., Grave, E., and Lample, G.
\newblock Llama: Open and efficient foundation language models, 2023{\natexlab{b}}.
\newblock URL \url{https://arxiv.org/abs/2302.13971}.

\bibitem[Vaswani(2017)]{vaswani2017attention}
Vaswani, A.
\newblock Attention is all you need.
\newblock \emph{Advances in Neural Information Processing Systems}, 2017.

\bibitem[Wolf et~al.(2020)Wolf, Debut, Sanh, Chaumond, Delangue, Moi, Cistac, Rault, Louf, Funtowicz, Davison, Shleifer, von Platen, Ma, Jernite, Plu, Xu, Scao, Gugger, Drame, Lhoest, and Rush]{wolf-etal-2020-transformers}
Wolf, T., Debut, L., Sanh, V., Chaumond, J., Delangue, C., Moi, A., Cistac, P., Rault, T., Louf, R., Funtowicz, M., Davison, J., Shleifer, S., von Platen, P., Ma, C., Jernite, Y., Plu, J., Xu, C., Scao, T.~L., Gugger, S., Drame, M., Lhoest, Q., and Rush, A.~M.
\newblock Transformers: State-of-the-art natural language processing.
\newblock In \emph{Proceedings of the 2020 Conference on Empirical Methods in Natural Language Processing: System Demonstrations}, pp.\  38--45, Online, October 2020. Association for Computational Linguistics.
\newblock URL \url{https://www.aclweb.org/anthology/2020.emnlp-demos.6}.

\bibitem[Ye et~al.(2021)Ye, Liu, Hu, Zhu, Yang, and Wang]{ye2021jslim}
Ye, R., Liu, L., Hu, S., Zhu, F., Yang, J., and Wang, F.
\newblock Jslim: Reducing the known vulnerabilities of javascript application by debloating.
\newblock In \emph{International Symposium on Emerging Information Security and Applications}, pp.\  128--143. Springer, 2021.

\bibitem[Zhang \& Ali-Eldin(2025)Zhang and Ali-Eldin]{zhang2025hiddenbloatmachinelearning}
Zhang, H. and Ali-Eldin, A.
\newblock {G}it{H}ub - jzh18/negativa --- github.com.
\newblock \url{https://github.com/negativa-ai/negativa/tree/main}, 2025.

\bibitem[Zhang et~al.(2023)Zhang, Alhanahnah, and Ali-Eldin]{zhang2023blafsbloatawarefile}
Zhang, H., Alhanahnah, M., and Ali-Eldin, A.
\newblock Blafs: A bloat aware file system, 2023.
\newblock URL \url{https://arxiv.org/abs/2305.04641}.

\bibitem[Zhang et~al.(2024)Zhang, Alhanahnah, Ahmed, Fatih, Leitner, and Ali-Eldin]{zhang2024machine}
Zhang, H., Alhanahnah, M., Ahmed, F.~A., Fatih, D., Leitner, P., and Ali-Eldin, A.
\newblock Machine learning systems are bloated and vulnerable.
\newblock \emph{Proceedings of the ACM on Measurement and Analysis of Computing Systems}, 8\penalty0 (1):\penalty0 1--30, 2024.

\bibitem[Ziegler et~al.(2019)Ziegler, Geus, Heinloth, H{\"o}nig, and Lohmann]{ziegler2019honey}
Ziegler, A., Geus, J., Heinloth, B., H{\"o}nig, T., and Lohmann, D.
\newblock Honey, i shrunk the elfs: Lightweight binary tailoring of shared libraries.
\newblock \emph{ACM Transactions on Embedded Computing Systems (TECS)}, 18\penalty0 (5s):\penalty0 1--23, 2019.

\end{thebibliography}
\bibliographystyle{mlsys2025}

\appendix
\newpage

\section{Appendix}
\begin{table*}[]
    \centering
    \caption{Jaccard Similarity of used functions and kernels in \texttt{tensorflow\_cc.so} used each pair of workloads. The bottom left shows the similarity of kernels between each pair of workloads. The top right shows the similarity of functions between each pair of workloads.}
    \label{tab:tf-similarity}
    \begin{small}
        \begin{tabular}{lllrrrr}
            \toprule
                                                  &                             &           & \multicolumn{2}{c}{\texttt{MobileNetV2}} & \multicolumn{2}{c}{\texttt{Transformer}}                                                        \\
                                                  &                             &           & \multicolumn{2}{c}{TensorFlow}           & \multicolumn{2}{c}{TensorFlow}                                                                  \\
            \cmidrule(r){4-5} \cmidrule(r){6-7}
                                                  &                             &           & Train                                    & Inference                                & Train                     & Inference                \\ \midrule
            \multirow{2}{*}{\texttt{MobileNetV2}} & \multirow{2}{*}{TensorFlow} & Train     & -                                        & 0.89  \cellcolor{gray!25}                & 0.95  \cellcolor{gray!25} & 0.89 \cellcolor{gray!25} \\
                                                  &                             & Inference & 0.5  \cellcolor{gray!54}                 & -                                        & 0.86 \cellcolor{gray!25}  & 0.82 \cellcolor{gray!25} \\ \hline
            \multirow{2}{*}{\texttt{Transformer}} & \multirow{2}{*}{TensorFlow} & Train     & 0.38  \cellcolor{gray!54}                & 0.29 \cellcolor{gray!54}                 & -                         & 0.88 \cellcolor{gray!25} \\
                                                  &                             & Inference & 0.02    \cellcolor{gray!54}              & 0.03 \cellcolor{gray!54}                 & 0.05  \cellcolor{gray!54} & -                        \\ \bottomrule
        \end{tabular}
    \end{small}
\end{table*}

\begin{table*}[hp]
    \caption{Debloating results of vLLM and Transformers with the top 9 LLMs using distributed inference.  }
    \label{tab:general_results}
    \begin{small}
        \begin{tabular}{llrrrrrrr}
            \toprule
            \multirow{2}{*}{Framework}             & \multirow{2}{*}{Model}            & \multirow{2}{*}{\#Lib.}  & \multirow{2}{*}{\makecell{Total File\\ Size/MB}} & \multicolumn{2}{c}{CPU Code} & \multicolumn{2}{c}{GPU Code}                           \\
                                 &                         &  &          & Size/MB                            & \#Functions                         & Size/MB    & \#Elements \\
            \midrule
            \multirow{9}{*}{vLLM}         & c4ai\_command\_r\_plus       & 137                & 3,790 (54)        & 650 (70)                           & 837K (93)                           & 1,910 (84) & 7,587 (85) \\
                                          & internlm2\_5\_7b\_chat       & 135                & 3,790 (54)        & 650 (70)                           & 837K (93)                           & 1,910 (84) & 7,587 (85) \\
                                          & llama\_3\_70b\_instruct      & 135                & 3,790 (54)        & 650 (70)                           & 837K (93)                           & 1,910 (84) & 7,587 (85) \\
                                          & mixtral\_8x22b\_instruct     & 137                & 3,790 (54)        & 650 (70)                           & 837K (93)                           & 1,910 (84) & 7,587 (85) \\
                                          & phi\_3\_medium\_4k\_instruct & 136                & 4,164 (54)        & 655 (71)                           & 848K (93)                           & 2,244 (79) & 8,297 (84) \\
                                          & qwen\_72b\_instruct          & 136                & 3,792 (54)        & 651 (70)                           & 837K (93)                           & 1,910 (84) & 7,587 (85) \\
                                          & qwen15\_110b\_chat           & 136                & 3,792 (54)        & 651 (70)                           & 837K (93)                           & 1,910 (84) & 7,587 (85) \\
                                          & yi\_15\_34b                  & 135                & 3,790 (54)        & 650 (70)                           & 837K (93)                           & 1,910 (84) & 7,587 (85) \\
                                          & zephyr\_orpo\_141b\_a35b     & 137                & 3,790 (54)        & 650 (70)                           & 837K (93)                           & 1,910 (84) & 7,587 (85) \\

            \midrule
            \multirow{9}{*}{Transformers} & c4ai\_command\_r\_plus       & 94                 & 2,866 (61)        & 514 (73)                           & 590K (92)                           & 1,592 (86) & 7,165 (85) \\
                                          & internlm2\_5\_7b\_chat       & 94                 & 2,866 (61)        & 514 (73)                           & 591K (92)                           & 1,592 (86) & 7,165 (85) \\
                                          & llama\_3\_70b\_instruct      & 96                 & 2,863 (61)        & 513 (73)                           & 587K (92)                           & 1,592 (86) & 7,165 (85) \\
                                          & mixtral\_8x22b\_instruct     & 98                 & 2,866 (61)        & 514 (73)                           & 591K (92)                           & 1,592 (86) & 7,165 (85) \\
                                          & phi\_3\_medium\_4k\_instruct & 94                 & 2,866 (61)        & 514 (73)                           & 590K (92)                           & 1,592 (86) & 7,165 (85) \\
                                          & qwen\_72b\_instruct          & 97                 & 2,866 (61)        & 513 (73)                           & 588K (92)                           & 1,592 (86) & 7,165 (85) \\
                                          & qwen15\_110b\_chat           & 98                 & 2,868 (61)        & 514 (73)                           & 590K (92)                           & 1,592 (86) & 7,165 (85) \\
                                          & yi\_15\_34b                  & 97                 & 2,866 (61)        & 514 (73)                           & 590K (92)                           & 1,592 (86) & 7,165 (85) \\
                                          & zephyr\_orpo\_141b\_a35b     & 97                 & 2,866 (61)        & 514 (73)                           & 590K (92)                           & 1,592 (86) & 7,165 (85) \\

            \bottomrule
        \end{tabular}
    \end{small}
\end{table*}

\begin{lstlisting}[language=asm,caption={Command line used for Nsys profiling},label={lst:nsys}]
nsys profile --trace=cuda -o report {Command to run the ML workload}
\end{lstlisting}

\end{document}